\documentclass{article}

\usepackage[final, nonatbib]{neurips_2020}

\usepackage{caption}

\usepackage[utf8]{inputenc} 
\usepackage[T1]{fontenc}    
\usepackage{hyperref}       
\usepackage{url}            
\usepackage{booktabs}       
\usepackage{amsfonts}       
\usepackage{nicefrac}       
\usepackage{microtype}      
\usepackage{graphicx}
\usepackage{xcolor}

\newcommand{\MYhref}[3][blue]{\href{#2}{\color{#1}{#3}}}%

\usepackage{titlesec}
\titlespacing*{\section}{0pt}{.1\baselineskip}{\baselineskip}
\titlespacing*{\subsection}{0pt}{1.0\baselineskip}{\baselineskip}

\title{Learnings from Frontier Development Lab and SpaceML - AI Accelerators for NASA and ESA}

\author{%
  Siddha Ganju  \thanks{Frontier Development Lab} \thanks{ SpaceML}  \\
  Nvidia Corporation\\
  \texttt{sganju@nvidia.com} \\
  \and
  
     \textbf{Anirudh Koul} \footnotemark[1] \footnotemark[2] \\
    Pinterest \\
  \texttt{anirudhkoul@gmail.com}\\
  \and

 \textbf{Alexander Lavin} \footnotemark[1] \\
 Latent Sciences \\
\texttt{lavin@latentsci.com} \\ 
 \and 
 
  \textbf{Josh Veitch-Michaelis} \footnotemark[1]  \\
  Liverpool John Moores University \\ 
 \texttt{j.veitchmichaelis@gmail.com}
  \and

     \textbf{Meher Kasam} \footnotemark[1] \footnotemark[2] \\
     Square \\
    \texttt{meherkasam@gmail.com}\\
    \and
  
  \textbf{James Parr} \footnotemark[1] \footnotemark[2] \\
  Trillium Technologies \\
  \texttt{james@frontierdevelopment.org}

}

\begin{document}

\maketitle


\begin{abstract}
Research with AI/ML technologies lives in a variety of settings with often asynchronous goals and timelines: academic labs and government organizations pursue open-ended research focusing on discoveries with long-term value, while research in industry is driven by commercial pursuits and hence focuses on short-term timelines and return on investment. 
The journey from research to product is often tacit or ad hoc, resulting in technology transition failures, further exacerbated when R\&D is interorganizational and interdisciplinary. 
Even more, much of the ability to produce results remains locked in the private repositories and know-how of the individual researcher, slowing the impact on future research by others and contributing to the ML community's challenges in reproducibility.
With research organizations focused on an exploding array of fields, opportunities for the handover and maturation of interdisciplinary research reduce. 
With these tensions, we see an emerging need to measure the correctness, impact, and relevance of research during its development to enable better collaboration, improved reproducibility, faster progress, and more trusted outcomes.
We perform a case study of the Frontier Development Lab (FDL), an AI accelerator under a public-private partnership from NASA and ESA. FDL research follows principled practices that are grounded in responsible development, conduct, and dissemination of AI research, enabling FDL to churn successful interdisciplinary and interorganizational research projects, measured through NASA's Technology Readiness Levels. We also take a look at the SpaceML Open Source Research Program, which helps accelerate and transition FDL's research to deployable projects with wide spread adoption amongst citizen scientists. 
\end{abstract}

\section{Introduction}
The potential of ML to truly be integrated into scientific decision making may be severely inhibited by transitional failures: key moments of breakdown in the evolution of complex AI methods and pipelines which is also a contributing factor to AI's reproducibility problem~\cite{conf_aaai_GundersenK18}.
The integrity of AI in science applications therefore needs a quality control system which can form a foundation for effective handover. This requirement is reflected in NASA Science Mission Directorate's (SMD) commitment to Open Science~\cite{smd}, which directs NASA Science to place emphasis on developing and implementing specific Open Science capabilities: continuous evolution of data and computing systems, better engagement of the scientific community and in turn strategic partnerships for innovation, and provide data access to the wider research community for robust validation of published research results. However, without a quality control system, this vision of Open Science stalls - bogged down by divergent research goals, heterogeneous data types, poorly annotated code and idiosyncratic nomenclature.

Several crowdsourcing platforms like Zooniverse, SETI@Home, LHC@Home, Amazon Mechanical Turk offer financial incentives or volunteer computing projects to involve citizen scientists. Global competitions (Kaggle, ImageNet, Alexa Prize) that offer datasets and compute have catalyzed interest in various fields of machine learning and inducted cohorts of students and citizen scientists. Additionally, challenges and hackathons like NASA International Space Apps Challenge aim to motivate and provide projects and resources to students. Such avenues offer a better on boarding experience for researchers and citizen scientists who are part of these platforms. The opportunities can be made more accessible through distributed open-source research that builds on leveraging existing products, like cutting edge enhanced data products and space AI research while avoiding the GitHub graveyard (unusable code and research). Additionally, such distributed open source research performed in accordance with the NASA Technology Readiness Levels (TRL)~\cite{Nasa2003NASASE} helps to classify the maturity of projects from research through production. The Frontier Development Lab (FDL) and SpaceML are such distributed opportunities that seek to involve citizen scientists at scale while enhancing educational and onboarding opportunities.

\section{Frontier Development Lab and SpaceML}

The Frontier Development Lab (FDL) is an international AI accelerator for both NASA and the European Space Agency (ESA), with additional corporate and government partners such as Google, Nvidia, the Mayo Clinic, and the United States Geological Survey (USGS), and the SETI Institute.
FDL serves as a useful sandbox for observing interdisciplinary ML and open science in real-time, over a broad range of science subjects.
FDL's mission is to apply AI technologies and scientific data towards solving some of the biggest challenges humanity faces, ranging from planetary defense~\cite{Zoghbi2017SearchingFL, JENNISKENS201821}, climate change~\cite{zantedeschi2020cumulo}, space traffic management~\cite{simoes2019fdl}, predicting solar weather~\cite{Szenicereaaw6548, Galvez_2019}, supporting disaster response~\cite{mateo2019flood}, astronaut health~\cite{antoniadou2019}, exploring and mapping the Moon~\cite{Moseley_2020}, detecting exoplanets~\cite{Ansdell_2018}, and astrobiology~\cite{2018LPI....49.1275C, Cobb_2019}.

To continue the accelerated momentum of projects beyond the research phase, eligible FDL projects are migrated to the ``SpaceML'' Open Source Research Program. SpaceML's first iteration in 2020, converted many high school students to citizen scientists with unique stories, like a student from Nigeria, who improved FDL’s advanced AI powered meteor detector pipeline \cite{imc_cams} along with an interactive interface in his local cyber-cafe.

FDL bridges the gap between industry, government, and academic research labs through its public-private partnership model, {democratizing opportunities globally}. Also bridging the gap between traditional aerospace projects and more recent software development projects, FDL leverages learnings from startup accelerators: {fast feedback loops, peer and expert reviews, parallel exploration and pivots}, and {fail-fast-succeed-sooner mindset}. Along with space engineering processes, such as {gated technology reviews}, FDL is nimble to build technologies rapidly, while keeping robustness and safety in check.

\begin{figure}[ht]
    \centering
    \includegraphics[width=\textwidth]{"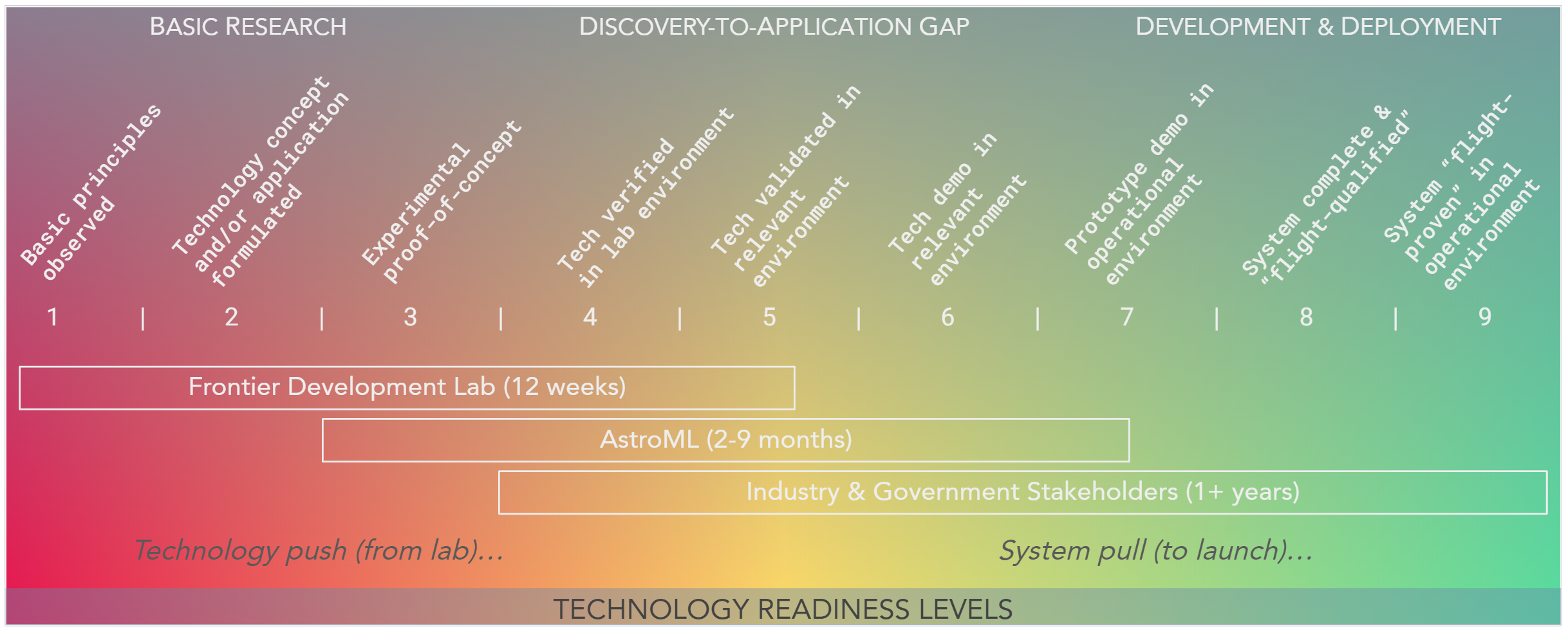"}
    \caption{
    Technology Readiness Levels for a space project or mission (as defined by ESA), where the operation areas of FDL, SpaceML, and space stakeholders roughly correspond to TRL 1-5, 3-7, and 4-9, respectively. 
    The TRL for a space system includes its submodules (e.g., flight software, comms, attitude control hardware, cameras, etc. in an Earth-observation satellite) which may include various AI models and algorithms. 
    FDL develops ML technologies that can later be integrated into larger software and space systems. To successfully transition R\&D to partner organizations (including SpaceML open-source community), it is critical that the delivered ML technologies are robust and reliable, as discussed in this paper. 
    FDL leverages the TRL4ML process, which defines an ML systems engineering framework that is motivated by the above space TRL, but tailored for ML applications  \cite{lavin2020technology}.
    }
    \label{trl-timeline}
\end{figure}

Effective research management through {iterative development}, {risk mitigation}, and {measuring quantifiable progress} are baked into the research cycle through peer reviewing and TRL4ML~\cite{lavin2020technology}. In this paper we use the term ``TRL'' or ``level'' to signify the exact stage of AI development in the TRL4ML framework as shown in Fig.\ref{trl-timeline}. Utilizing the TRL4ML framework helps FDL and SpaceML practice responsible development and dissemination of interdisciplinary AI and sciences research, enabling stakeholders to get multi-year grants to pursue the projects to deployment.
TRL4ML helps FDL to address NASA SMD's goals of enabling open science with robust, well-tested workflows. Even more, FDL is a committed leader in ML reproducibility and open-source software.


Our case study focuses on FDL and SpaceML practices leading to responsible development, conduct, and dissemination of interdisciplinary AI and sciences research allowing stakeholders to get multi-year grants to pursue the project to deployment. In particular, we look at the FDL research cycle and measure out best practices that have served FDL well. Our case study is organized as follows, we first take a look at how FDL and SpaceML are organized, their process flow, early indicators of success and its five-year impact.

\begin{figure}[ht]
    \centering
    \includegraphics[width=\textwidth]{"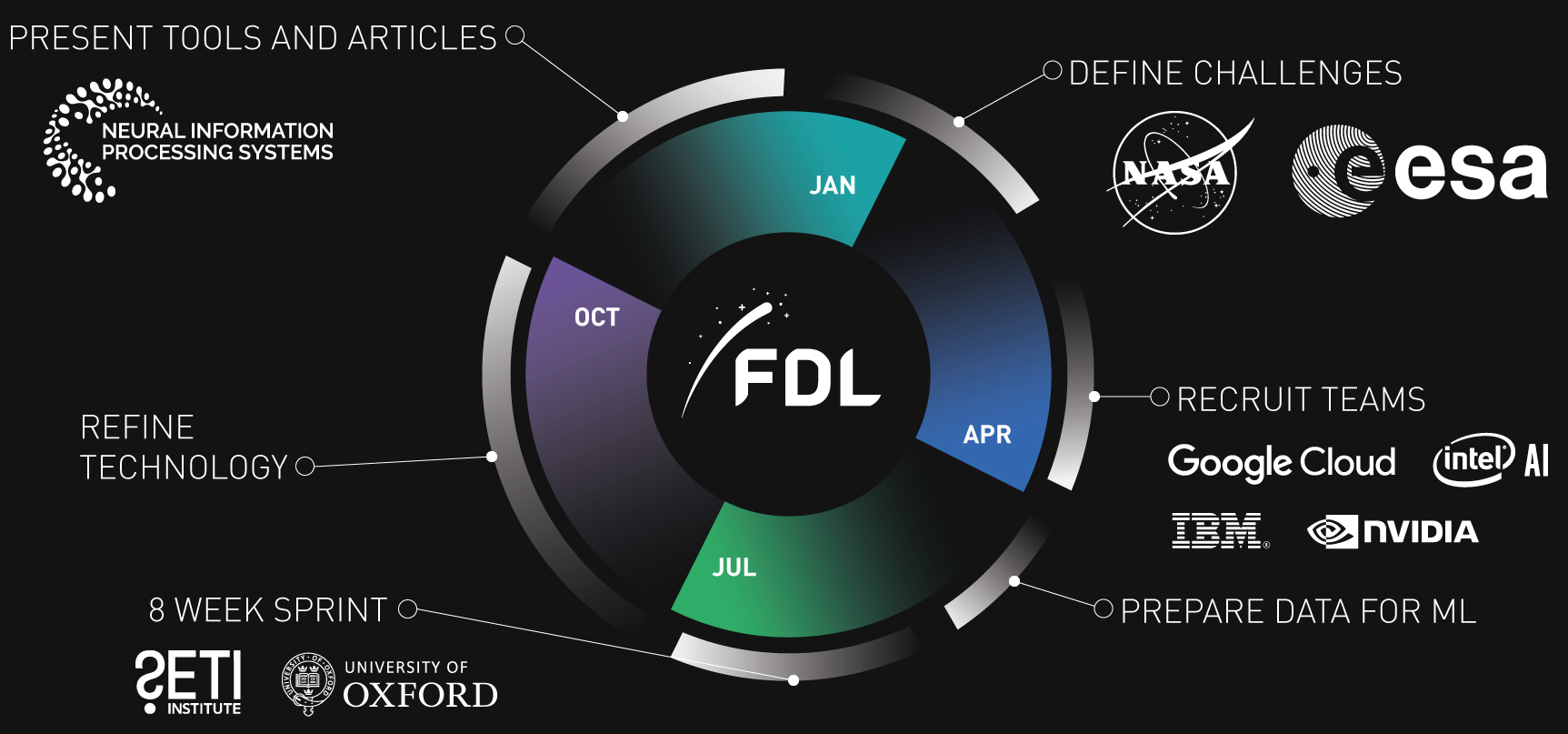"}
    \caption{Even though FDL researchers gather and focus on a task for an twelve-week research sprint, the work happens all year round from inviting challenges to gathering compute and data to having subject matter experts and stakeholders on board.  }
    \label{fdl-timeline}
\end{figure}

\begin{table}[ht]
\caption{Demography distribution in FDL\\
Legend: (M)ale, (F)emale, (P)ublished, (D)eployed AI, Proceedings link embedded behind years\\}
\label{tab:demography}
\resizebox{\textwidth}{!}{%
\begin{tabular}{|l|l|l|l|l|l|l|}
\hline
\textbf{FDL year} &
  \textbf{\# Participants} &
  \textbf{\% Gender} &
  \textbf{\% Ethnic Minority} &
  \textbf{FDL Location} &
  \textbf{Number of projects} &
  \textbf{Proceedings} \\ \hline
\textbf{\MYhref{https://en.calameo.com/read/0055032804c2759a5c960?authid=wb8VPRbf3ugF}{2016}} &
  12 &
  \begin{tabular}[c]{@{}l@{}}M: 67\\ F: 33\end{tabular} &
  41 &
  USA &
  3 &
  \begin{tabular}[c]{@{}l@{}}P:2\\ D:1\end{tabular} \\ \hline
\textbf{\MYhref{https://en.calameo.com/read/0055032804c2759a5c960?authid=wb8VPRbf3ugF}{2017}} &
  24 &
  \begin{tabular}[c]{@{}l@{}}M: 71\\ F: 29\end{tabular} &
  17 &
  USA &
  5 &
  \begin{tabular}[c]{@{}l@{}}P:3\\ D:2\end{tabular} \\ \hline
\textbf{\MYhref{https://en.calameo.com/read/005503280c171307d20fb?page=1}{2018}} &
  28 &
  \begin{tabular}[c]{@{}l@{}}M: 86\\ F: 14\end{tabular} &
  32 &
  USA, Europe &
  \begin{tabular}[c]{@{}l@{}}USA = 6\\ EU = 2\end{tabular} &
  \begin{tabular}[c]{@{}l@{}}P:9\\ D:3\end{tabular} \\ \hline
\textbf{\MYhref{https://en.calameo.com/read/0048123632374547b7087?page=1}{2019}} &
  36 &
  \begin{tabular}[c]{@{}l@{}}M: 67\\ F: 33\end{tabular} &
  17 &
  USA, Europe &
  \begin{tabular}[c]{@{}l@{}}USA = 6\\ EU = 3\end{tabular} &
  \begin{tabular}[c]{@{}l@{}}P:11\\ D:3\end{tabular} \\ \hline
\textbf{2020} &
  45 &
  \begin{tabular}[c]{@{}l@{}}M: 71\\ F: 29\end{tabular} &
  33 &
  USA Europe &
  \begin{tabular}[c]{@{}l@{}}USA = 8\\ EU = 3\end{tabular} &
  \begin{tabular}[c]{@{}l@{}}P:30\\ D:WIP\end{tabular} \\
  \hline
\end{tabular}%
}
\end{table}

\subsection{FDL Timeline}
\label{subsec:fdltimeline}
While the core research at FDL is conducted over a twelve-week period (a sprint), the groundwork extends throughout the year as shown in Fig \ref{fdl-timeline}. Here we discuss a detailed timeline of FDL and SpaceML.

\textbf{Partner/Stakeholder Selection - T-6 months}: Project stakeholders (e.g. different teams at NASA, USGS, ESA) are selected who can provide the high impact focus area for investigation, budget for researchers \& mentors, expertise, and data. Commercial sponsors also join in, often also offering compute (e.g. Google Cloud, Intel, NVIDIA), data (e.g. satellite imagery from Planet Labs), and expertise. SETI Institute hosts FDL at their office location in Mountain View, California, apart from handling organizational and administrative tasks.

\textbf{Challenge Selection - T-5 months}: The Challenge Selection stage, primarily centered around a global public meeting called the “Big Think” is focused on which challenge or task can be included in the sprint. Attended by over 200 subject matter experts from academia, industry, and humanitarian organizations in diverse fields, the Big Think is free for the public to attend and comment along the following lines: (1) Why are these challenges pressing? (2) What impact does an AI/ML solution have on the challenges? (3) How can you contribute?. Initial ideas by the stakeholders are discussed and further brainstormed by subgroups on approaches, feasibility, risk, impact, scope, finally transforming the initial focus area into an actionable project plan. Pre-read information about the focus areas is made available prior to the event, allowing the attendees to prepare, maximizing the outcome of the event. The range of fields represented has often resulted in discovering potential flaws which without the inter-disciplinary diversity in fields would not have been possible. Post-event, the attendees submit project-specific feedback, as well as people in their connection networks who can accelerate the project’s success.

Often FDL projects are among the first AI forays in a topic area, thus it is beneficial to have cross-functional teams with the ability to identify appropriate tools and ideas from both ML and science domains. Initiatives are challenging and success is largely uncertain, so utilizing early feedback to iterate on goals and pivoting when necessary are important.

\textbf{Team Selection - T-3 months}: Post the ‘Big Think’ event, a global call for both researchers and mentors interested in participating in these summer challenges is made. 
A team consists of 7-8 members: (1) four researchers, one with deep AI expertise, one with an advanced science competency, and two ‘hybrid-skilled’ researchers with backgrounds in both AI and sciences, (2) team mentors - one or two science experts, and an ML expert (3) a “Super Mentor” who has hybrid expertise in both the science domain and ML and has previously mentored a successful team. Thus the FDL format blends the expertise and capacity of academia and commercial partners to support rapid experimentation in building ML models in data intensive areas. The special sauce is its interdisciplinary research teams, which are composed of subject specialists from the space sciences and specialists from the data sciences at the Ph.D. or post doc level. While open to citizens of any country, stakeholders may mandate different team criteria - for example, NASA funded projects require 50\% representation of US citizens. Table~\ref{tab:demography}  details the yearly demographic distribution.

\textbf{Data Preparation - T-2 weeks}: To hit the ground running, the mentors work with the stakeholders and partners for several weeks to ensure that ready to use data is available. This includes any license procurement, loading the data to cloud storage (usually being in tens to hundreds of gigabytes), API access, and scripts to preprocess data. Some datasets or hardware may require a “request for access” such as medical data from astronauts, and data from remote observatories, or a hardware simulator (such as the Lunar Landing simulator), and a drone (for detecting fallen meteors), etc has to be done before the team selection.

\textbf{Bootcamp - T-1 week}: The Bootcamp is the official launch of FDL every summer. The focus is on introductions and developing a working rapport with different members of the team. To ensure the same foundational skills, this week incorporates practical tutorials on AI methods, collaboration tools, version control, distributed processing, coding practices, cloud usage and includes lightning talks with experts to showcase the state of art in AI. The team members also work on a mini-project, called "The Big Why" (described in the next section) to help develop interpersonal relationships. All these learning exercises help people from different backgrounds (AI and sciences) get on the same page and be ready for the high paced research environment. 

\textbf{The Big WHY - T-1 week}: The bootcamp culminates with a presentation from each team answering `The Big Why' - understanding the vision behind their challenge, potential impact, and how it relates to real life. This is an exploratory phase that usually results in interesting questions, introspection, and detailed discussions with the experts in the FDL network and stakeholders. The value of this exercise for the researchers is to internalize the vision and feel motivated to solve it.

\textbf{Goal setting and weekly sprints - T+0-8 week}: 
Teams are given freedom to choose how to approach their challenge, defining their goals according to three grades - safe, stretch, and moonshot. The team jointly decides and iterates on the feasibility of the goals during the first week. Over the course of the sprint, teams may push the goalposts further from safe to stretch and sometimes even beyond the moonshot if time allows. FDL sprints are essentially an Agile~\cite{beck2001manifesto} software development process applied to research. At the end of each week, teams are required to present their progress to a panel of peers, stakeholders and external experts, providing regular and gated feedback and critique. The emphasis on rapid prototyping and parallel exploration helps teams validate approaches and mitigate risks early and often. 

\textbf{Showcase}: The end deliverables are twofold: (1) presentations for project sponsors and other stakeholders (summarized into a NASA Technical Memorandum), which teams may adapt for formal publications in journals or conference venues 
(2) software deliverables as open-source packages for continued development by stakeholders and/or others in the SpaceML community. FDL, then, proactively invites volunteer open-source contributors under the {SpaceML Open Source Research Program} 
to take the project closer to the final vision. Thus, SpaceML opens research opportunities, previously only available to people with advanced academic backgrounds, and grow citizen scientists by motivating them to play a role in high value research. 
Teams also work with a speaking and delivery coach to deliver a final 8-minute TED-style presentation which are streamed live publicly and have a global audience. The storytelling format makes highly technical content accessible to the general public, while still providing enough details to excite scientists who can then read the team’s detailed report.  

\subsection{SpaceML Timeline}
\label{subsec:SpaceMLtimeline}

\textbf{Challenge}: FDL researchers create a set of miniature extension challenges on a scale of easy to hard. Most challenges are a super-focused topic developable by the competent developer to a prototype level in a 3-week full-time period. The challenge definition includes details of technical stack, reference research papers, and potential solution ideas. 
Once the contributions reach some maturity, new volunteer mentors through the FDL network join in to level-up the project further.

\textbf{Development}: To enable ease of getting started, FDL researchers package code and instructions on GitHub, Google Colab (with free GPUs) so that the new contributors can run it with relative ease. The GitHub repository also contains a video explaining the overall vision of each project and descriptions of project-specific beginner challenges, along with detailed guidance on how to get started. The beginner challenges, akin to an interview, allow competent and motivated contributors to move further in the challenge. At the same time, for people who are not yet familiar with the technical stack (science or AI), this serves as an educational opportunity to level up with the help of the detailed guides. Beginner guides can include ``How to get started in AI'', ``How to use GitHub effectively'', ``How to access satellite imagery data'', ``Code Quality standards'', ``Citation guide'', and, ``Code Reproducibility standards''. Contributors who have shown enough progress are then provided with Google Cloud credits to scale up their AI training and storage requirements. Contributors work directly with the stakeholders to define timelines for development and the deliverables to advance TRL levels.

\textbf{Mentorship}: FDL researchers, who worked on the original project, are now the stakeholders of the open-source project, and switch their roles to a mentor while providing weekly mentorship. Throughout the process of improving the project, the contributors eventually become an expert in the niche area of research. Once the contributions reach some maturity, new mentors through the FDL network join in to level up the project further. For example, an experienced industry engineer can perform code review, while an experienced scientist can review the experiments for the correctness and suggest other ways to amplify the value of the contributions. SpaceML provides access to experienced and often well-known professionals as volunteer mentors to the public who might not be accessible otherwise. To make efficient use of time for both the mentors and mentees, the mentees need to earn the time of mentors in the form of completed work, like through the early challenge or the weekly goals.

\textbf{Showcase}: SpaceML volunteers regularly present their progress to stakeholders and deliberate ways to increase TRL level and the optimum method to convert the research into production. SpaceML focuses on higher TRL projects and as such each showcase session is detailed with information about the complete pipeline ranging from data discovery, ingestion and processing to model training, optimization and deployment. SpaceML has streamlined a series of showcase events, the ``AI Affinity Networks'' where the contributors, stakeholders and future users of the project discuss the best way to make the work of the contributors available for public access.

\subsection{Early Indicators of Success}
Year over year, FDL learns new practices to disseminate research and discover potential research impacts early on, making future FDL iterations even more successful. 
The combination of curated challenges, close mentorship, community of expertise and an emphasis on rapid prototyping has ensured a high success rate for FDL and SpaceML. Practices that have stood out as key components contributing to success include: stakeholder involvement, weekly peer and expert reviewing, mentorship, and TRL4ML. We utilize the example of the NASA CAMS~\cite{JENNISKENS201140} project to exemplify  the utility of these practices.\footnote{The CAMS project, established in California in 2010, uses hundreds of low light CCTV cameras to capture the meteor activity in the night sky. After each night, the resident scientist performed the triangulation of tracklets captured by two or more cameras and computed the meteor's trajectory, orbit, and lightcurve. Each solution was manually classified as a meteor or not a meteor (i.e., planes, birds, clouds etc). CAMS became a part of FDL in 2017 in order to automate the data processing pipeline and filtration of meteors.}

\paragraph{Stakeholder Involvement.} The process of bringing ML research into production is non-trivial, in particular when a myriad of teams and organizations are involved.
Interorganizational and interdisciplinary R\&D takes this a step further, where the involvement of stakeholders who often act as decision- and policy-makers is critical. Throughout the process, they are able to amplify benefits and mitigate potential risks while involving AI and science researchers. 
More importantly, when these R\&D opportunities are open to the public in a structured way, it attracts and grows more citizen scientists from diverse backgrounds who previously might not have the chance to participate due to socioeconomic or educational constraints. 
FDL and SpaceML promote a culture of responsible openness that focuses on communication, quantifiable progress, reproducibility, and knowledge-sharing. Although not necessarily research in their strictest forms, these components are important in the research life-cycle, and, have helped accelerate deployment of socially responsible AI products. As a direct result of the progress of the CAMS AI project and the widespread public interest it generated, the stakeholders in turn received funding leading to a 4x expansion in the number of camera stations including Chile, Namibia, and New Zealand with the establishment of multiple CAMS stations in the southern hemisphere.

\paragraph{Peer and Expert Reviewing.} Frequent reviews tie FDL and SpaceML research to TRL4ML progress definitions. Teams work towards standardized, weekly formal review sessions where they showcase work and seek feedback. The review panel alternates weekly between internal FDL peers and external experts. 
The review processes systematically reduces technical debt in the following ways:
\textbf{(1) Regular, weekly, reviews} ensure that teams are held accountable, but also allows for early identification of potential issues and suggestions from peer review. By soliciting weekly feedback on the incremental work, the teams are able to identify what works and what does not work, increasing the chances of success while ensuring alignment with the ultimate goal 
\textbf{(2) Gated reviews} are a key aspect of FDL through the inclusion of stakeholders and domain experts. Reviewers have the power to change the course of a project if the team's approach appears to be infeasible or inappropriate, potentially saving considerable amounts of wasted time. Based on the work and quantifiable results presented by the team, the expert reviewers and stakeholders can decide the path forward, or circle back with quantifiable improvements
\textbf{(3) Actionable feedback} from reviewers helps guide teams to using appropriate methods to solve their problems. FDL has a close relationship with many industrial stakeholders which also enables teams to seek help on software engineering aspects of their projects.
For the CAMS project, user interface (UI) or dashboards were created as a mock up first, verified through A/B tests with the experts and then developed and deployed 
\textbf{(4) Quantifiable metrics for end-user objectives} are important for showing improvement over both established ML baselines and in-use methods. Here, the inclusion of domain experts is critical as many FDL projects involve collating novel datasets and evaluating prior research. Stress is on the significance of the numerical improvements and comparison to existing baselines. For CAMS - the night sky activity is largely governed by non-meteors, but the scientists are interested in meteors, hence focus on using metrics that mitigated data imbalance were used after consultation with the experts.

\paragraph{Mentorship.} {Distributed open-source research} provides to each volunteer contributor (1) a theme of interest to learn and develop e.g. AI or space sciences (2) a support structure of diverse experts and mentors (3) an end-to-end plan that details how their work builds into the final product. Contributors are motivated by research with high public value that has a path towards deployment. High school and early graduate school students from a wide variety of socioeconomic backgrounds and geographies volunteered in the first iteration of SpaceML.

\paragraph{TRL4ML.} Leveraging TRL4ML to define technology maturity in a quantitative way enables FDL and SpaceML teams to develop robust AI technologies and communicate readiness to stakeholders and project owners. 
TRL4ML emphasizes regular checkpointing that pushes ML development towards interoperability, reliability, maintainability, extensibility, and scalability. Additionally, there are a number of components that FDL and SpaceML specifically find valuable: 
\textbf{(1) Risk mitigation}, specifically the identification and testing of various failure modes; as the project matures, it's important to take into account any areas the ML model(s) could potentially fail (sometimes even silently), and develop fail-safes when integrating the model(s) into a larger solution pipeline.
\textbf{(2) Flexibility and modularity to scale} as demands arise, which is non-trivial when working with dataflows that imply nondeterministic model outputs. As the CAMS project grew globally, requirements to add features to the UI arose. Features being worked on during SpaceML 2020 include a search mechanism to search by meteor shower name, ability to view movements in yearly meteor showers, and the ability to view several nights of observations together
\textbf{(3) Monitoring and updating} for both performance metrics (e.g. latency, and accuracy for model drift) \textit{and} data shifts. With higher TRLs the ability to regularly update data and models is an integral part of the pipelines, for example, weather changes, smoke, and cloud patterns affect the view of the night sky such that corresponding data validation checks were incorporated into the CAMS AI pipeline.
\textbf{(4) Data bias identification and mitigation} to ensure the data is representative; for example, selection bias due to feedback loops or limited annotators may be common, and these must be proactively checked throughout the project lifecycle.
\textbf{(5) Automated testing and delivery}, such as CI/CD pipelines that can trigger tests automatically, to run ML- and data-specific stress and smoke tests. Manual intervention can be triggered when needed, such as sending low confidence meteors for verification to scientists in the CAMS project.  For the CAMS project, the AI pipeline is automated and runs regular tests, production of benchmarks, and sends low confidence meteors for verification to scientists.
\textbf{(6) Documentation of collective wisdom} by way of cross-functional teams and reviews, open-source code repositories\footnote{FDL and SpaceML release publicly accessible code on Astrorepo: \MYhref{https://github.com/FrontierDevelopmentLab}{github.com/FrontierDevelopmentLab}}, and wikis for knowledge sharing. Synthetic data, models or web tools produced as a result of FDL such as the code behind CAMS UI is also released.
\textbf{(7) Code quality and software engineering standards} are important to maintain for robust systems, but often at odds with how researchers operate; FDL and SpaceML leans on industry mentors to lead with sound software practices.

An important part of TRL4ML evident in both FDL and SpaceML is that the development paths are often {non-monotonic}. That is, an ML technology can reach proof-of-concept status (level 3-4) and then revert to a previous level for further validations or algorithmic changes. FDL and SpaceML handles this well in several capacities, for example iterating on project paths over successive years, and even within the weekly sprints as project pivots. Non-monotonic development by-default helps ensure robust, reliable AI technologies are eventually shipped.

\subsection{Impact}
Originally starting as a NASA AI Accelerator in 2015, the success of FDL led investments from stakeholders beyond NASA, culminating in a more internationally collaborative research environment and leading to the launch of FDL Europe funded by ESA, and challenges supported in Canada and Asia-Pacific through stakeholders like the USGS, ESA, Canada Space Agency, Luxembourg Space Agency, Australian Space Agency to name a few. 

With a total of eight iterations featuring 37 challenges over five years, FDL \& FDL Europe have 70+ publications at premier AI and science journals, including {Science Advances, Planetary Science Journal, Acta Astronautica} and conferences such as {AAAI, ICML, NeurIPS} and many others. FDL projects have been featured by {TechCrunch, Fast Company, NASA, Forbes}, and multiple AI documentaries including “Age of AI” (hosted by Robert Downey Jr., 45M+ views). FDL work has also received the {Best Paper Award at NeurIPS Climate Change and AI 2019 workshop}~\cite{zantedeschi2020cumulo}. Five FDL projects have been deployed - or are in the process of deployment and 12 tools and enhanced datasets are planned to be released. FDL through its global admission has {28\% female participation} vs 22\% industry average, and {28.8\% ethnic minority representation}. See Table \ref{tab:demography} for more details. 

AstroRepo acts as the \textit{Open Research Portal} through which FDL and SpaceML provide easily browsable research accompanied by open source code and a portfolio of starter projects that newcomers can play around and learn from. The open source contributors are able to experiment within the AstroRepo provided sandbox such as through Google Colab (with free GPUs) so that the contributors can run the code and pipelines with relative ease. AstroRepo also tries do do away with esoteric nomenclature and strives to make the information accessible and easy to comprehend. These projects range from understanding data pipelines to training new models on the existing annotated data. The existing projects available on AstroRepo include preprocessed data, implemented algorithms, trained models, and end to end pipelines that detail the life cycle of research and products. Thus FDL and SpaceML inspire narratives of space exploration and understanding of our planet through interdisciplinary AI and science projects. 

The first iteration of SpaceML was held in Fall 2020 on developing a remote sensing search system for {\MYhref{https://earthdata.nasa.gov/esds/impact}{NASA IMPACT}}. While the original FDL project consisted of post-docs and PhDs, the majority of SpaceML contributors were high school students, with participants from Nigeria, Mexico, Korea, Germany, and USA. The first cohort of students started out with no prior AI background knowledge yet produced results on problems ranging from self-supervised learning using SimCLR, multi-resolution image search, cost-efficient data labeling, balancing imbalanced data without labels, and generating synthetic patches for missing satellite imagery. 
Other contributors integrated it into~\MYhref{https://worldview.earthdata.nasa.gov}{NASA Worldview} for browsing global satellite images. In 4 months, SpaceML helped convert a TRL-3 project to TRL-6, providing stakeholders with evidence to grow funding, and helping them move towards the ultimate goal -- searching NASA’s 33 PB of unlabelled satellite imagery in seconds (versus the current multi-week timeline) thereby significantly improving productivity of research scientists. 

Another project in SpaceML's repertoire was the  TRL-9 NASA CAMS project. For the CAMS project (part of FDL 2017) - beyond the huge efficiency improvements from automation, an AI pipeline with human-level accuracy, and a new web tool that allowed global access to the previous night's activity, FDL teams contribution resulted in raising awareness and bringing in new citizen scientists who established stations in Brazil. The automation of the data processing allowed researchers to manage the camera stations better, which led to the {detection of the highest number of meteors in a single night (including 3003 Geminids \& 1154 sporadic meteors) in December 2017}. In 2020, the quick turnaround time of discovery from the CAMS web tool ultimately led to the an almost monthly reporting of a new and unusual meteor showers in the sky, including the {discovery of multiple meteor showers}. One of those showers helped better define the orbit of parent comet Grigg-Mellish, which was observed poorly in 1907~\cite{JENNISKENS2020104979}. During SpaceML several features like a search mechanism to search by meteor shower name, ability to view movements in yearly meteor showers, and the ability to view several nights of observations together, and enhancing and updating the UI or web tool. 

\section{Retrospective}
FDL projects aim to enhance existing workflows with AI and through its publications and results has been successful in developing a functional prototype that may be applicable to a real-world task, but, not all projects are converted to real-world applications. There may be multiple reasons behind this - time constraints (8-weeks), limited human resources available to see the project throughout its lifetime, and many others. 

Currently, applications are more prominent from countries that either already has a space agency or have local organizations that are able to fund researchers and domain leads in an FDL sprint. Researchers and domain leads often find out about FDL from their networks and social media. There is a definite need to increase participation from missed out countries.

\section{Conclusion}
We shared a case study on AI R\&D practices that enable NASA's Frontier Development Lab and SpaceML to consistently churn out ambitious, successful projects. Specifically several mechanisms are instrumental: 
\textbf{(1) Stakeholder involvement} through decision- and policy-makers is critical. They are able to amplify benefits and mitigate potential risks while involving AI and science researchers, and opening opportunities for citizen scientists. 
\textbf{(2) Regular peer and expert reviews} provide actionable feedback and serve as key decision points to measure the correctness, impact, and relevance of research. 
\textbf{(3) Mentorship} when provided in a structured way can attract and grow citizen scientists from diverse backgrounds who previously might not have the chance to participate due to socioeconomic or educational constraints. 
\textbf{(4) TRL4ML} quantifies progress and formulates communication via a common technical language (especially for interdisciplinary R\&D), enabling better collaboration, improved reproducibility, faster progress, and more trusted outcomes. TRL4ML provides a guideline to build responsible ML, utilizing existing best practices from systems engineering, coupled with considerations unique to ML. 

Ultimately these techniques aim to \textbf{prioritize open research, responsible publication, and development of reliable and user-centered ML systems}. Even more, by coordinating scientific ML efforts across industry, academia, and government, FDL and SpaceML promote a culture of responsible openness helping accelerate deployment of ML projects with socially responsible AI products.

\bibliography{references}
\bibliographystyle{IEEEtran}

\end{document}